\documentclass[
amssymb,nobibnotes,aps,pre,showpacs]{revtex4}
\usepackage{graphics}
\usepackage[dvipdfm]{graphicx}
\usepackage{bmpsize}
\usepackage{graphicx}
\usepackage{color}
\usepackage{subfigure}
\usepackage{amsmath}
\usepackage{amssymb}
\usepackage{bm}
\usepackage{float}
\usepackage{epsfig}
\usepackage{wrapfig}
\usepackage{epstopdf}
\usepackage{lmodern}
\usepackage{color}
\usepackage[utf8]{inputenc}

\newcommand\be{\begin{equation}}
\newcommand\e{\end{equation}}
\newcommand\ba{\begin{eqnarray}}
\newcommand\ay{\end{eqnarray}}

\begin{document}

\title{{ Effective photon mass and exact \\ translating quantum relativistic structures}}

\author{Fernando Haas}
\email{fernando.haas@ufrgs.br}
\affiliation{Physics Institute, Federal University of Rio Grande do Sul,  Avenida Bento Gon\c{c}alves 9500, CEP 91501-970, Porto Alegre, RS, Brazil}

\author{Marcos Antonio Albarracin Manrique}
\email{sagret10@hotmail.com}
\affiliation{Physics Institute, Federal University of Rio Grande do Sul,  Avenida Bento Gon\c{c}alves 9500, CEP 91501-970, Porto Alegre, RS, Brazil}

\begin{abstract}
\noindent
Using a variation of the celebrated Volkov solution, 
the Klein-Gordon equation for a charged particle  
is reduced to a set of ordinary differential equations, exactly solvable in specific cases. The new quantum relativistic structures can reveal a localization in the radial direction perpendicular to the wave packet propagation, thanks to a non-vanishing scalar potential. The external electromagnetic field, the particle current density and the charge density are determined. The stability analysis of the solutions is performed by means of numerical simulations. 
The results are useful for the description of a charged quantum test particle in the relativistic regime,  provided spin effects are not decisive.
\end{abstract}


\maketitle

\section{Introduction}
The analysis of systems in a very high energy density needs the consideration of both quantum and relativistic effects.
This is certainly true in extreme astrophysical environments like white dwarfs and neutron stars, where the de Broglie length is comparable to
the average inter-particle distance, making quantum diffraction effects appreciable, and where temperatures reach relativistic levels. In addition, the development of strong X-ray free-electron lasers \cite{Hand} allows new routes for the exploration of matter on the angstrom scale, where quantum effects are prominent, together with a quiver motion comparable to the rest energy. Optical laser intensities of $10^{25} \,{\rm W/cm}^2$, and above, are expected to trigger radiation-reaction effects in the electron dynamics, allowing to probe the structure of the quantum vacuum, together with copious particle-antiparticle creation \cite{diPiazza}. We are entering a new era to test fundamental aspects of light and matter interaction in extreme limits. In particular there is the achievement of a continuous decrease of laser pulse duration accompanied by the increase of the laser peak intensity \cite{Mourou}, motivating the detailed analysis of fundamental quantum systems under strong electromagnetic (EM) fields. The interaction of such strong EM fields with solid or gaseous targets is expected \cite{Wang1} to create superdense plasmas of a typical density up to $10^{34} \,{\rm m}^{-3}$. For instance, the free-electron laser Linac Coherent Light 
Source (LCLS) considers powerful femtosecond coherent soft and hard X-ray sources operating on wavelengths as small as $0.06 \,{\rm nm}$, many orders of magnitude smaller than the conventional lasers systems acting on the micrometer scale \cite{Ishikawa}. The nonlinear collective photon interactions and vacuum polarization in plasmas \cite{Marklund},  the experimental assessment of the Unruh effect \cite{Unruh, Crispino}, and of the linear and nonlinear aspects of relativistic quantum plasmas \cite{book}, are fruitful avenues of fundamental research. Moreover, there is a renewed interest on quantum relativistic-like models related to graphene \cite{Novoselov}, narrow-gap semiconductors and topological insulators \cite{Zawadzki1}.

In this work we investigate the quantum relativistic dynamics of a test charge.  
Since typical test charges are electrons and positrons (fermions), a complete treatment would involve the Dirac equation. However, for processes where the spin polarization is not decisive, a possible modeling can be based on the Klein-Gordon equation (KGE). The adoption of the KGE is a valid approximation in view of the analytical complexity of the Dirac equation, especially if a strong magnetization is not present. For instance, the QED cascade process, which provides diverse tests of basic predictions of QED and theoretical limits on achievable laser intensities, is known to be not strongly spin-dependent \cite{Ehlotzky}. Naturally, the scalar particle approach excludes problems like the collapse-and-revival spin dynamics of strongly laser-driven electrons \cite{Skoromnik} or the Kapitza-Dirac effect \cite{Kapitza, Freimund}, where the spin polarization is essential. The analysis of spin effects will be left for a forthcoming communication.

Recently, there has been much interest on KGE based models. Examples are provided by the analysis of the {\it Zitterbewegung} (trembling motion) of Klein-Gordon particles in extremely small spatial scales, and its simulation by classical systems \cite{Rusin}, the KGE as a model for the Weibel instability in relativistic quantum plasmas \cite{Mendonca5}, the description of standing EM solitons in degenerate relativistic plasmas \cite{Mikaberidze}, the KGE as the starting point for the wave kinetics of relativistic quantum plasmas \cite{Mendonca4}, the KGE in the presence of a strong rotating electric field and the QED cascade \cite{Raicher3}, the Klein-Gordon-Maxwell multistream model for quantum plasmas \cite{Haas1}, the negative energy waves and quantum relativistic Buneman instabilities \cite{Haas2}, the separation of variables of the KGE in a curved space-time in open cosmological universes \cite{Villalba}, the resolution of the KGE equation in the presence of Kratzer \cite{Saad} and Coulomb-type \cite{Bakke} potentials, the KGE with a short-range separable potential and interacting with an intense plane-wave EM field \cite{Faizal}, electrostatic one-dimensional propagating nonlinear structures and pseudo-relativistic effects on solitons in quantum semiconductor plasma \cite{Wang2},  the square-root KGE \cite{Haas3}, hot nonlinear quantum mechanics \cite{Mahajan}, a quantum-mechanical free-electron laser model based on the single electron KGE \cite{Yan},  and the inverse bremsstrahlung in relativistic quantum plasmas \cite{Mendonca2}.

Very often, the treatment of charged particle dynamics described by the Klein-Gordon or Dirac equations assumes a circularly polarized electromagnetic (CPEM) wave \cite{Mendonca2}-\cite{Cronstrom}, mainly due to the analytical simplicity. However, the CPEM wave is not the ideal candidate for particle confinement. It is the main purpose of the present work to pursue an alternative route, where a perpendicular compression is realized in terms of appropriate scalar and vector potentials. We investigate the possibility of relatively simple EM field configurations for which exact solutions localized in a transverse plane are available, therefore providing new benchmark structures for the KGE. For this purpose the wave function will be described by a modified Volkov {\it Ansatz} \cite{Volkov}, incorporating an extra transverse dependence as explained in Sec. \ref{esol}. Separability of the KGE is then obtained for appropriate EM field configurations. 

Unlike in a vacuum, in ionized media the self-consistent EM field is analog to a massive  field, where the corresponding effective photon mass is obtained from the plasma dispersion relation \cite{Mendonca3, Akhiezer}. Already in 1953, Anderson \cite{Anderson} has observed the formal analogy between the wave equations for the scalar and vector potentials in ionized media, and the evolution equations for a massive vector field. Shortly after this has motivated the concept of massive Higgs boson \cite{Higgs}. In order to achieve the development of the new exact solutions, 
the appearance of an effective photon mass $m_{\rm ph}$ in a plasma will be decisive. Observe that the photon mass in this case is an effective one, not a ``true'' photon mass as proposed in alternative theories. The ``real'' value of the photon mass was experimentally estimated  \cite{Tu} to be as small as $10^{-49} \,{\rm kg}$, several orders of magnitude smaller than the effective photon mass in a typical ionized medium.

The paper is organized in the following way. In Sec. II, the modified Volkov {\it Ansatz} is introduced, and the EM fields compatible with it are determined, so that the KGE becomes separable. The resulting structures are shown to be dependent on the specific form of the scalar potential, entering as the main input in the determining equation for the radial wave function. In Sec. III, this determining equation is solved in concrete cases. In this way the oscillatory compressed test charge density is explicitly derived. Sec. IV considers in more detail the physical parameters relevant for the problem, from extremely dense plasmas arising in laser-plasma compression experiments to astrophysical compact objects such as white dwarfs. The conservation laws of total charge and energy are derived, and used to verify the numerical methods applied to check the stability of the exact solution against perturbations. Sec. IV presents some conclusions. 

\section{Exact solution}
\label{esol}
We shall consider the problem of a charged scalar particle (charge $q$, mass $M$) coupled to the EM four-potential $A_\mu = (\phi/c, {\bf A})$. The metric tensor will be taken as
$g^{\mu\nu} = {\rm diag}(1,-1,-1,-1)$, so that with a photon four-wave-vector $k_\mu = (k_0 = \omega/c, {\bf k})$ in the laboratory frame and with $x_\mu = (x_0 =  c\,t, {\bf r})$ one has, e.g., the four-product $k \cdot x = k^{\mu} x_{\mu} = k_0 x_0 - {\bf k}\cdot{\bf r}$, with the summation convention implied.
In this setting and using the minimal coupling assumption, the covariant form of the KGE reads
\begin{equation}
\label{a1}
(P^\mu - q A^\mu) (P_\mu - q A_\mu) \Psi = M^2 c^2 \Psi \,,
\end{equation}
where $P_\mu = \Bigl((i\hbar/c)\partial/\partial t, - i\hbar\nabla\Bigr)$ is the four-momentum operator and $\Psi$ is the complex charged scalar field. Considering the Lorentz gauge
\begin{equation}
\label{lordao}
\partial_\mu A^\mu = (1/c^2)\,\partial\phi/\partial t + \nabla\cdot{\bf A} = 0 \,,
\end{equation}
using $\partial_\mu = \partial/\partial x^\mu$, a more explicit form of the KGE is
\begin{equation}
\label{a2}
\hbar^2 \Box\Psi + 2i\hbar q\left(\frac{\phi}{c^2} \frac{\partial\Psi}{\partial t} + {\bf A}\cdot\nabla\Psi\right) - q^2\left(\frac{\phi^2}{c^2} - |{\bf A}|^2\right) \Psi + M^2 c^2 \Psi = 0 \,,
\end{equation}
where $\Box = (1/c^2)\,\partial^2/\partial t^2 - \nabla^2$ is the d'Alembertian operator.

A brief examination of the literature will be shown to be suggestive. Numerous works \cite{Mendonca2}-\cite{Cronstrom} on the KGE assume 
a (right-handed) circularly polarized electromagnetic (CPEM) wave. For a monochromatic field with four-wave-vector 
$k_\mu = (\omega/c, 0, 0, k)$, it amounts to
\begin{equation}
\label{a3}
{\bf A} = \frac{A_0}{\sqrt{2}} ({\boldsymbol \epsilon}\, e^{i\theta} + {\boldsymbol \epsilon}^* e^{-i\theta}) \,, \quad \phi = 0 \,,
\end{equation}
where $A_0$ is a slowly varying function of the phase
\begin{equation}
\label{phase}
\theta = k \cdot x = \omega t - k z \,,
\end{equation}
while ${\boldsymbol \epsilon} = (\hat{x} - i \hat{y})/\sqrt{2}$ denotes the polarization vector, with the unit vectors $\hat{x}, \hat{y}$ perpendicular to the direction of light propagation. The motivation for the CPEM assumption is due to practical reasons, since it can be most easily implemented in laser experiments, as well as to formal reasons, due to the reduction of the quantum wave equation to a well-known ordinary differential equation, namely a Mathieu equation \cite{Becker, Cronstrom, Polyanin}.
In the case of a Dirac field in vacuum, a similar procedure allows the construction of the celebrated Volkov solution \cite{Volkov}, provided the four-vector potential depends on the phase only.

In the present work, a radically different avenue is chosen. Instead of assuming {\it ab initio} a CPEM wave, the EM field is left undefined as far as possible, requiring the KGE to be still reducible to certain ordinary differential equations (to be specified later). Nevertheless, most of the usual steps toward the Volkov solution are maintained. As will be proved, a large class of field configurations will be so determined. The results put the Volkov solution into a perspective, and considerably enlarge the class of fields for which benchmark analytic results in a quantum relativistic plasma can be accessible in principle.

In a similar spirit of the derivation of the Volkov solution \cite{Volkov}, it is now assumed
\begin{equation}
\label{a4}
\Psi = \exp\left(-\,\frac{i p \cdot x}{\hbar}\right)\,\psi({\bf r}_\perp,\theta) \,,
\end{equation}
where $p_\mu = ({\cal E}/c, {\bf p})$ is the constant asymptotic four-momentum of the particle, far from the EM field. The mass-shell condition $p^\mu p_\mu = ({\cal E}/c)^2 - |{\bf p}|^2 = M^2 c^2$ holds throughout.  Moreover, the transverse dispersion relation
\begin{equation}
\label{a5}
k^\mu k_\mu = \frac{\omega^2}{c^2} - k^2 = \frac{m_{\rm ph}^2 c^2}{\hbar^2} \,,
\end{equation}
is supposed, where $m_{\rm ph}$ is the effective photon mass acquired due to screening in the plasma \cite{Akhiezer}. The photon mass can be self-consistently calculated using quantum electrodynamics \cite{Raicher2} but here will be considered mostly as an input data. Unlike Volkov's solution, a dependence of the envelope wave function on transverse coordinates is allowed in Eq. (\ref{a4}), where for light propagation in the $z-$direction one has $\hat{z}\cdot{\bf r}_\perp = 0$. As a matter of fact, the extra transverse dependence is found to be crucial in what follows. The { direction of propagation of the wave packet reflected in the proposed wave function} breaks the isotropy. 
Although the relation between $\omega$ and $k$ could be left completely undefined, 
the transverse plasma dispersion relation is assumed to keep resemblance with the previous analysis in the literature \cite{Mendonca2}--\cite{Akhiezer}.

Substitution of the {\it Ansatz} (\ref{a4}) into the KGE, taking into account the mass-shell condition and the dispersion
relation (\ref{a5}), gives
\begin{eqnarray}
- \hbar^2 \nabla_{\perp}^2 \psi &+& m_{\rm ph}^2 c^2 \frac{\partial^2 \psi}{\partial\theta^2} + 2 i\hbar \Bigl(q{\bf A} - {\bf p}\Bigr)\cdot\nabla_{\perp}\psi
+ 2 i\hbar \left[\frac{\omega}{c^2} \Bigl(q\phi - {\cal E}\Bigr) - k\Bigl(q A_z - p_z\Bigr)\right]\frac{\partial\psi}{\partial\theta} \nonumber \\
&+&  \left[\Bigl|q {\bf A} - {\bf p}\Bigr|^2 - \frac{1}{c^2}\Bigl(q \phi - {\cal E}\Bigr)^2\right]\psi + M^2 c^2 \psi = 0 \,, \label{KGEe}
\end{eqnarray}
where $\nabla_\perp = \hat{x}\,\partial/\partial x + \hat{y}\,\partial/\partial y$ and ${\bf A} = (A_x, A_y, A_z)$.

For the sake of reference, in the case of the CPEM field (\ref{a3}), assuming $\nabla_\perp \psi = 0$, and defining
\begin{equation}
\tilde{\psi} = \exp\left[- \, \frac{i\hbar}{m_{\rm ph}^2 c^2}\Bigl(\frac{\omega {\cal E}}{c^2} - k p_z\Bigr)\theta\right] \psi \,,
\end{equation}
the result \cite{Becker, Cronstrom} from Eq. (\ref{KGEe}) is the Mathieu equation \cite{Polyanin},
\begin{equation}
\frac{d^2\tilde{\psi}}{d\tilde{\theta}^2} + \frac{4}{m_{\rm ph}^2 c^2} \left[q^2 A_{0}^2 + \frac{\hbar^2}{m_{\rm ph}^2 c^2}\Bigl(\frac{\omega {\cal E}}{c^2} - k p_z\Bigr)^2 - 2 q A_0 p_\perp \cos(2\tilde{\theta})\right]\tilde{\psi} = 0 \,,
\end{equation}
where
\begin{equation}
\tilde{\theta} = \frac{1}{2}(\theta - \theta_0) \,, \quad \tan\theta_0 = \frac{p_y}{p_x} \,, \quad p_\perp = \sqrt{p_{x}^2 + p_{y}^2} \,.
\end{equation}

Back to the general case, and shifting the four-potential according to
\begin{equation}
\label{shi}
A_\mu \Rightarrow \tilde{A}_\mu =  A_\mu - p_\mu/q
\end{equation}
transforms the KGE (\ref{a2}) into 
\begin{equation}
\label{aa2}
\hbar^2 \Box\Psi + 2(q\tilde{A}^\mu+p^\mu)P_\mu \Psi - (q\tilde{A}^\mu + p^\mu)(q\tilde{A}_\mu + p_\mu)\Psi + M^2 c^2 \Psi = 0 \,,
\end{equation}
and Eq. (\ref{KGEe}) into
\begin{equation}
- \hbar^2 \nabla_{\perp}^2 \psi + m_{\rm ph}^2 c^2 \frac{\partial^2 \psi}{\partial\theta^2} + 2 i\hbar q\tilde{\bf A}\cdot\nabla_{\perp}\psi + 2 i\hbar q\left(\frac{\omega\tilde{\phi}}{c^2}
- k\tilde{A}_z\right)\frac{\partial\psi}{\partial\theta} + q^2 \left(|\tilde{\bf A}|^2 - \frac{\tilde\phi^2}{c^2}\right)\psi + M^2 c^2 \psi = 0 \,, \label{KGE}
\end{equation}
the later equation does not exhibiting the asymptotic four-momentum $p_\mu$. In what follow, the tilde symbol over the four-potential will be omitted, for simplicity. Notice that the Lorentz gauge is still attended by the displaced four-potential.

Instead of sticking to the search of pure traveling wave solutions as usually done, we want to investigate the possibility of localized wave-packets in the transverse plane also. This is a recommendable trend, having in mind (for instance) the usefulness of laser fields having a dependence on the transverse coordinates too, as in the case of focused beams. To keep some simplicity
consider solutions with a definite $z$ angular momentum component,
\begin{equation}
\label{psi}
\psi = \frac{e^{im\varphi}}{\sqrt{r}} R(r) S(\theta) \,,
\end{equation}
where the factor $1/\sqrt{r}$ was introduced just for convenience, $m = 0, \pm 1, \pm 2,...$ is the azimuthal quantum number and $(r,\varphi,z)$ are cylindrical coordinates, while $R, S$ are real functions to be determined. Naturally $L_z\,\psi \equiv - i\,\hbar\,\partial\psi/\partial\varphi = m\,\hbar\,\psi$.  Differently from twisted plasma waves \cite{Tito}, here the angular momentum is possibly carried by matter waves, not necessarily by EM waves.

Substituting the proposal (\ref{psi}) into Eq. (\ref{KGE}) gives
\begin{eqnarray}
- \frac{\hbar^2}{R}\frac{d^2 R}{dr^2} &+& \frac{m_{\rm ph}^2 c^2}{S} \frac{d^2 S}{d\theta^2} + M^2 c^2 + \frac{\hbar^2}{r^2}\left(m^2-\frac{1}{4}\right) + q^2 \left(|{\bf A}|^2 - \frac{\phi^2}{c^2}\right) - \frac{2 m \hbar q}{r} A_\varphi + \nonumber \\
&+& 2 i \hbar q \frac{\sqrt{r} A_r}{R} \frac{d}{dr}\left(\frac{R}{\sqrt{r}}\right)  + 2 i \hbar q \left(\frac{\omega \phi}{c^2} - k A_z\right) \frac{1}{S} \frac{dS}{d\theta} = 0 \,, \label{sep}
\end{eqnarray}
where $\phi = \phi(r,\theta), {\bf A} = A_{r}(r,\theta)\,\hat{r} + A_\varphi(r,\theta)\,\hat{\varphi} + A_z(r,\theta)\hat{z}$ with components supposed to be dependent on $(r, \theta)$ only, for consistency.

It is natural to seek for separable variables solutions. For this purpose, Eq. (\ref{sep}) must be the sum of parts individually containing either $r$ or $\theta$. Avoiding excessive constraints on $R, S$ at this stage,  from inspection of the terms proportional to $dR/dr$ or $dS/d\theta$, and since uninteresting solutions ($dS/d\theta = 0$ or $R \sim \sqrt{r}$) are ruled out, the following necessary conditions follow,
\begin{equation}
\label{ar}
A_r = \tilde{A}_{r}(r) \,, \quad A_z = \frac{\omega}{c^2 k}\phi + \tilde{A}_{z}(\theta) \,,
\end{equation}
where $\tilde{A}_{r}$ and $\tilde{A}_{z}$ must be functions of the indicated arguments. In this way the prescription of $R, S$ is postponed as long as possible.

More stringent conclusions follows since $\tilde{A}_{r}(r)$ does not contribute neither to $\bf{E}$ or $\bf{B}$.
In addition, inserting $A_z$ in the Lorentz gauge condition (\ref{lordao}) gives $d\tilde{A}_{z}(\theta)/d\theta = 0$, so that $\tilde{A}_z$ is a constant, with no contribution to the EM field also. Hence, without loss of generality it can be set
\begin{equation}
\label{arr}
\tilde{A}_r = \tilde{A}_z = 0 \,.
\end{equation}

Summing up the results until now, Eq. (\ref{sep}) becomes
\begin{equation}
- \frac{\hbar^2}{R}\frac{d^2 R}{dr^2}+ \frac{m_{\rm ph}^2 c^2}{S} \frac{d^2 S}{d\theta^2} + M^2 c^2 - \frac{\hbar^2}{4 r^2} + \left(q A_\varphi - \frac{\hbar m}{r}\right)^2 + \frac{m_{\rm ph}^2 q^2 \phi^2}{\hbar^2 k^2} = 0 \,.
\label{sepp}
\end{equation}

In principle, $A_\varphi$ and $\phi$ can be functions of $(r,\theta)$. However, it can be observed that for transverse EM fields the longitudinal 
components vanish so that
\begin{eqnarray}
E_z &=& - \frac{\partial\phi}{\partial z} - \frac{\partial A_z}{\partial t} = - \frac{m_{\rm ph}^2 c^2}{\hbar^2 k^2} \frac{\partial\phi}{\partial\theta} \equiv 0 \quad \Rightarrow \quad \phi = \phi(r) \,, \\
B_z &=& \frac{1}{r}\frac{\partial}{\partial r}\left(r A_\varphi\right) \equiv 0 \quad \Rightarrow \quad A_\varphi = \frac{F(\theta)}{r}  \,, \label{aphi}
\end{eqnarray}
Actually from Eq. (\ref{aphi}) one derives
\begin{equation}
\nabla\times\left(A_\varphi \, \hat\varphi\right) = \frac{k}{r}\frac{dF}{d\theta} \hat{r} + F(\theta) \delta^{2}({\bf r}_\perp)\,\hat{z} \,,
\end{equation}
where $\delta^{2}({\bf r}_\perp)$ is the two-dimensional delta function in the transverse plane, contributing a vortex line except if $F(\theta) = 0$. This choice will be adopted to avoid singularity at this stage, so that $A_\varphi = 0$.

Collecting results, we find
\begin{equation}
\label{4p}
\phi = \phi(r) \,, \quad {\bf A} = \frac{\omega}{c^2\,k}\,\phi(r)\,\hat{z}
\end{equation}
and the final form of the re-expressed KGE is
\begin{equation}
- \frac{\hbar^2}{R}\frac{d^2 R}{dr^2}+ \frac{m_{\rm ph}^2 c^2}{S} \frac{d^2 S}{d\theta^2} + M^2 c^2 + \frac{\hbar^2}{r^2}\left(m^2 - \frac{1}{4}\right) + \frac{m_{\rm ph}^2 q^2 \phi^{2}(r)}{\hbar^2 k^2} = 0 \,,
\label{seppp}
\end{equation}
which is obviously separable.

Denoting $P_{0}^2 > 0$ as the separation of variables constant, we get
\begin{eqnarray}
\label{s}
m_{\rm ph}^2 c^2 \frac{d^2 S}{d\theta^2} &+& P_{0}^2 S = 0 \,,\\
\label{r}
\hbar^2 \frac{d^2 R}{dr^2} &+& \left[P_{0}^2 - M^2 c^2 - \frac{\hbar^2}{r^2}\left(m^2 - \frac{1}{4}\right) - \frac{m_{\rm ph}^2 q^2 \phi^{2}}{\hbar^2 k^2}\right] R = 0 \,.
\end{eqnarray}
The requirement $P_{0}^2 > 0$ is adopted to avoid constant or unbounded solutions as $\theta \rightarrow \pm \infty$. It should be noted that the procedure makes sense only in a plasma medium
($m_{\rm ph} \neq 0$) to avoid triviality. Actually from the very beginning the limit $m_{\rm ph}/M \rightarrow 0$ changes the basic structure of the governing equations and should be treated as a singular perturbation problem \cite{Nayfeh}, as apparent from Eq. (\ref{KGEe}).

In specific calculations, like for calculations of cross sections, the non-shifted  four-potential is necessary. In view of Eq. (\ref{4p}), we would have the original scalar potential given by $Mc^2/q + \tilde{\phi}(r)$, and the original vector potential given by ${\bf p}/q + [\omega/(c^2 k)] \tilde{\phi}(r)\,\hat{z}$, where $\tilde{\phi}(r)$ is an arbitrary function of $r$ only.  In this way both the wavefunction given in Eq. (\ref{a4}) and the four-potential will contain the four-momentum.

 One can choose the origin of time so that $S(0) = 0$ so that from Eq. (\ref{s}) the longitudinal part of the wave function can be written as
\begin{equation}
\label{ss}
S(\theta) = \frac{1}{\sqrt{\pi}}\sin(n\theta) \,, \quad n = \frac{P_0}{m_{\rm ph}c} \,.
\end{equation}

To sum up, Eq. (\ref{a4}) represents an exact solution for the KGE for a charged scalar in the presence of a transverse plasma wave, provided the traveling envelope function $\psi$ in 
Eq. (\ref{psi}) is defined in terms of $R(r), S(\theta)$ satisfying the uncoupled linear system of second-order ordinary differential equations (\ref{s}) and (\ref{r}). The corresponding static EM field is
\begin{equation}
\label{emf}
{\bf E} = - \frac{d\phi}{dr} \hat{r} \,, \quad {\bf B} = - \frac{\omega}{c^2 k} \frac{d\phi}{dr} \hat{\varphi} \,,
\end{equation}
with a Poynting vector
\begin{equation}
\frac{1}{\mu_0} {\bf E}\times{\bf B} = \frac{\varepsilon_0 \,\omega}{k} \left(\frac{d\phi}{dr}\right)^2 \hat{z}
\end{equation}
along the wave propagation direction as expected, and an EM energy density
\begin{equation}
\frac{\varepsilon_0}{2}\,|{\bf E}|^2 + \frac{1}{2\,\mu_0}\,|{\bf B}|^2 = \varepsilon_0\,\left[1 + \frac{1}{2}\,\left(\frac{m_{\rm ph}\,c}{\hbar\,k}\right)^{2}\right]\,\left(\frac{d\phi}{dr}\right)^2 \,,
\end{equation}
where $\varepsilon_0, \mu_0$ are, respectively, the vacuum permittivity and permeability.
Notice the amplitude of the wave remains arbitrary, due to the linearity of the KGE.

For the sake of interpretation we can examine the conserved charged 4-current
\begin{equation}
\label{j}
J_\mu = \frac{q}{2M}\Bigl(\Psi^* (P_\mu - p_\mu - q A_\mu)\Psi + {\rm c.c.}\Bigr)
\end{equation}
associated to the particle, where c.c. denotes the complex conjugate. The extra term $\sim p_\mu$ in Eq. (\ref{j}) is needed in view of the shift (\ref{shi}). Writing $J_\mu = (c\rho, {\bf J})$ one derives
\begin{equation}
\label{den}
\rho = - \frac{q^2 \phi}{M c^2} |\Psi|^2 \,, \quad {\bf J} = \frac{q}{M}\left(\frac{m \hbar}{r}\, \hat\varphi - \frac{\omega\,q\,\phi}{c^2 k}\, \hat{z}\right)\,|\Psi|^2 \,,
\end{equation}
where $|\Psi|^2 = R^2 S^2/r$. As can be verified, indeed $\partial_\mu J^\mu = 0$ along solutions. From Eq. (\ref{den}) it is seen that the charge density $\rho$ associated to the test charge shows a radial dependence allowing for radial compression, together with an oscillatory pattern in the direction of wave propagation through $S(\theta)$. The density current ${\bf J}$ has a swirl provided $m \neq 0$, besides a longitudinal component.

We observe that the force density is
\begin{equation}
\label{ford}
\rho\,{\bf E} + {\bf J}\times{\bf B} = - \left(\frac{m_{\rm ph}\,c}{\hbar\,k}\right)^2\rho\,{\bf E} \,,
\end{equation}
opposite to the electric force density $\rho\,{\bf E}$, possibly implying a transverse confinement of the test charge, depending on the properties of the scalar potential. This radial confinement is certainly not possible in a vacuum, where the effective photon mass is exactly zero.

The charge density allows one to express the normalization condition as
\begin{equation}
\label{norma}
\int\,d{\bf r}\,\rho = q \quad \Rightarrow \quad \int_{0}^{\infty}\,\phi(r)\,R^{2}(r)\,dr = - \frac{M\,c^2}{q\,\Delta} \,,
\end{equation}
where $\Delta$ is the longitudinal extension of the system, or $z \in [- \Delta/2,\,\Delta/2]$.

The current density associated to the test charge should not be confused with the current density $J_{\mu}^{\rm ext} = (c\rho^{\rm ext}, {\bf J}^{\rm ext})$ responsible for the external EM field. One finds
\begin{equation}
\label{ext}
\rho^{\rm ext} = \varepsilon_0 \nabla\cdot{\bf E} = - \frac{\varepsilon_0}{r} \frac{d}{dr}\left(r \frac{d\phi}{dr}\right) \,, \quad
{\bf J}^{\rm ext} = \frac{1}{\mu_0} \left(\nabla\times{\bf B} - \frac{1}{c^2}\frac{\partial{\bf E}}{\partial t}\right) = \frac{\omega \rho^{\rm ext}}{k}\,\hat{z} \,,
\end{equation}
having a purely radial dependence, and a plasma flow in the longitudinal direction only, corresponding to a z-pinch configuration. 

Finally we present the field invariants
\begin{equation}
{\bf E}\cdot{\bf B} = 0 \,, \quad |{\bf E}|^2 - c^2 |{\bf B}|^2 = - \left(
\frac{m_{\rm ph}\,c}{\hbar\,k}\right)^2 \left(\frac{d{\phi}}{dr}\right)^2
\,.
\end{equation}

Although quite simple, the new explicit exact solution has not been officially recognized in the past, to the best of our knowledge. The reason perhaps is the need of an oscillating longitudinal part $S(\theta)$, which is possible only for a test charge in a plasma ($m_{\rm ph} \neq 0$). Moreover, the procedure has shown the solution to be the only one satisfying the following requirements: (a) extended Volkov {\it Ansatz} incorporating the transverse dependence, as shown in Eq. (\ref{a4}); (b) the dispersion relation (\ref{a5}); (c) separation of variables according to Eq. (\ref{psi}). In the following Section, illustrative examples are provided.

\section{Examples}

\subsection{Compressed structures}
\label{3a}
Following an inverse strategy, instead of first defining the scalar potential, for the sake of illustration we consider the radial function
\begin{equation}
R(r) = e^{-X/2}\,X^{|m|/2 + 1/4}\,\Phi(X) \,,
\label{rr}
\end{equation}
where $X = r^2/(2\,\sigma^2)$, $\sigma$ is an effective length and $\Phi = \Phi(X)$
satisfies Kummer's equation \cite{Polyanin},
\begin{equation}
\label{ku}
X\, \frac{d^2\Phi}{dX^2} + \left(1 + |m| - X\right)\,\frac{d\Phi}{dX} + \alpha\,\Phi = 0 \,.
\end{equation}
In addition, $\alpha$ is a parameter defined by
\begin{equation}
\label{alp}
\alpha = \frac{1}{2}\left((P_{0}^2 - M^2 c^2)\,\frac{\sigma^2}{\hbar^2} - 1 - |m| - \frac{1}{H^2}\right) \,,
\end{equation}
where
\begin{equation}
\label{dif}
H = \frac{\hbar^2\,k}{m_{\rm ph}\,|q\,\phi_0|\,\sigma} \,.
\end{equation}
is a dimensionless quantum diffraction parameter with $\phi_0 = \phi(0)$. Without loss of generality $k > 0$ is assumed.

The form (\ref{rr}) has recently attracted attention in the case of non-relativistic theta pinch quantum wires \cite{Kushwava}. Inserting
Eq. (\ref{rr}) into the radial equation (\ref{r}), taking into account Kummer's equation, and Eq. (\ref{alp}), we find the simple expression
\begin{equation}
\label{sca}
\phi =  \phi_0 \sqrt{1 + \frac{H^2 X}{2}}
\end{equation}
which according to Eq. (\ref{emf}) corresponds to
\begin{equation}
\label{eb}
{\bf E} = - \frac{\hbar^4\,k^2\,r\,\hat{r}}{4\,m_{\rm ph}^2\,\sigma^4\,q^2\,\phi} \,, \quad {\bf B} = - \frac{\hbar^4 \omega\,k\,r\,\hat{\varphi}}
{4\,m_{\rm ph}^2\,c^2\,\sigma^4\,q^2\,\phi} \,.
\end{equation}

The general solution to Eq. (\ref{ku}) is
\begin{equation}
\label{so}
\Phi = c_1 \,{\cal M}(-\alpha, 1+|m|, X) + c_2 \,{\cal U}(-\alpha, 1+|m|, X) \,,
\end{equation}
where $c_{1,2}$ are integration constants, ${\cal M}(-\alpha, 1+|m|, X)$ is the Kummer confluent hypergeometric function and ${\cal U}(-\alpha, 1+|m|, X)$ is the confluent hypergeometric function. Since ${\cal U}$ is always singular for $X \rightarrow 0$, we set $c_2 = 0$. Therefore, from Eq. (\ref{rr})and taking into account \cite{Polyanin} the asymptotic properties of ${\cal M}(-\alpha, 1+|m|, X)$, one has
\begin{equation}
\label{as}
R \sim \frac{\Gamma(1+|m|)}{\Gamma(-\alpha)}\,e^{X/2} X^{- \alpha - \frac{3}{4} - \frac{|m|}{2}}\,\Bigl(1 + {\cal O}(1/X)\Bigr) \,,
\end{equation}
where $\Gamma$ is the gamma function. In addition, $R$ is well-behaved at the origin, with $R(0) = 0$.

In view of Eq. (\ref{as}), it follows that the solution is unbounded for large $X$, unless the infinite series defining the Kummer confluent hypergeometric function terminates. It is apparent that this happens if and only if $1/\Gamma(-\alpha) = 0$, implying $\alpha = l = 0, 1, 2,...$. In this case ${\cal M}(- l, 1+|m|, X)$ becomes proportional to a Laguerre polynomial. Hence we derive the quantization condition
\begin{equation}
\label{qua}
P_{0}^2 = M^2 c^2 + \frac{\hbar^2}{\sigma^2}\left(1 + |m| + 2 l + \frac{1}{H^2}\right) > M^2 c^2\,.
\end{equation}
Since $P_0 = n m_{\rm ph} c$ [see Eq. (\ref{ss})], and in view of the small value of the photon mass, in general a large $n$ is necessary to fulfill Eq. (\ref{qua}).

\begin{figure}
\includegraphics[angle=0,scale=0.4]{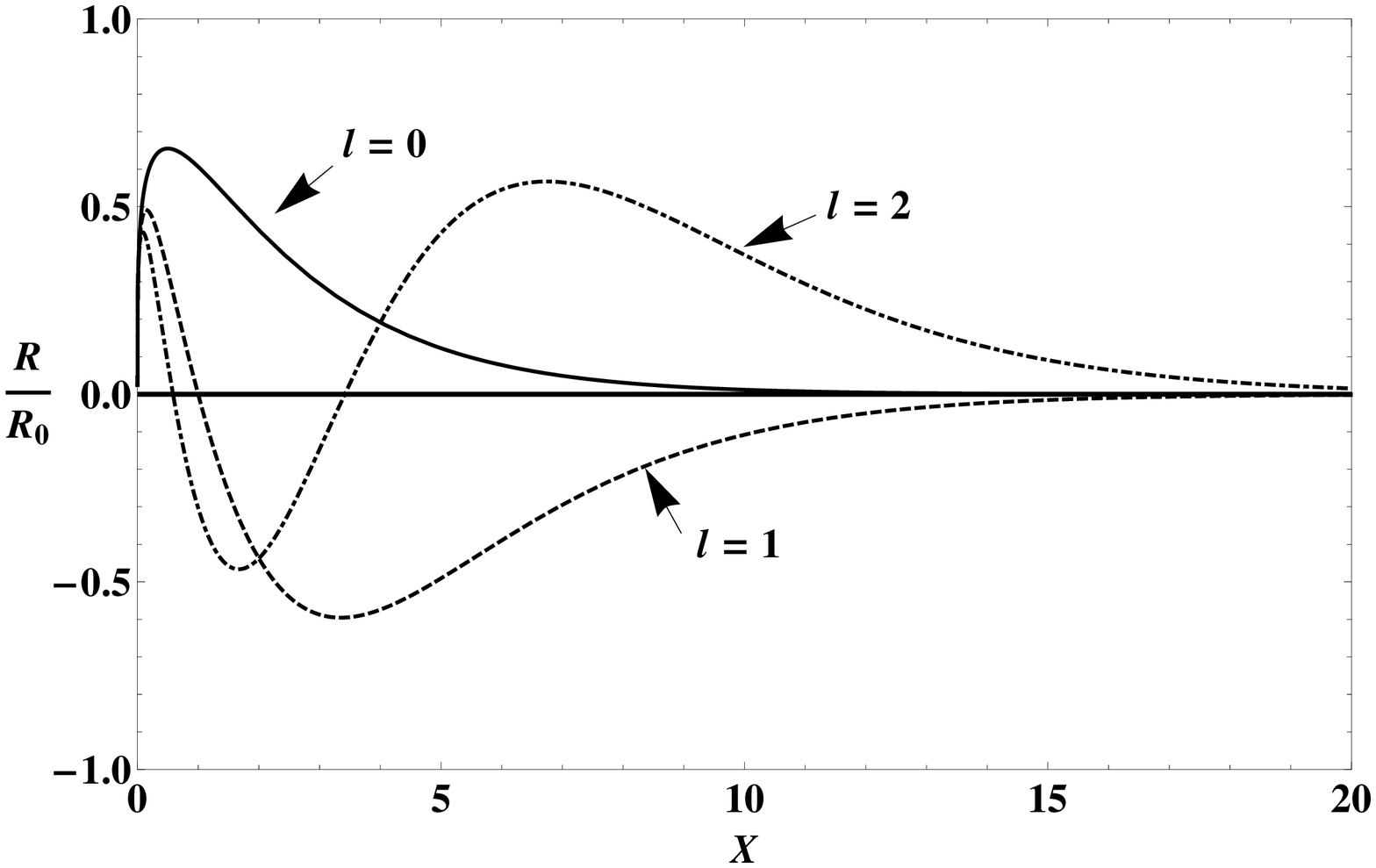}
\includegraphics[angle=0,scale=0.4]{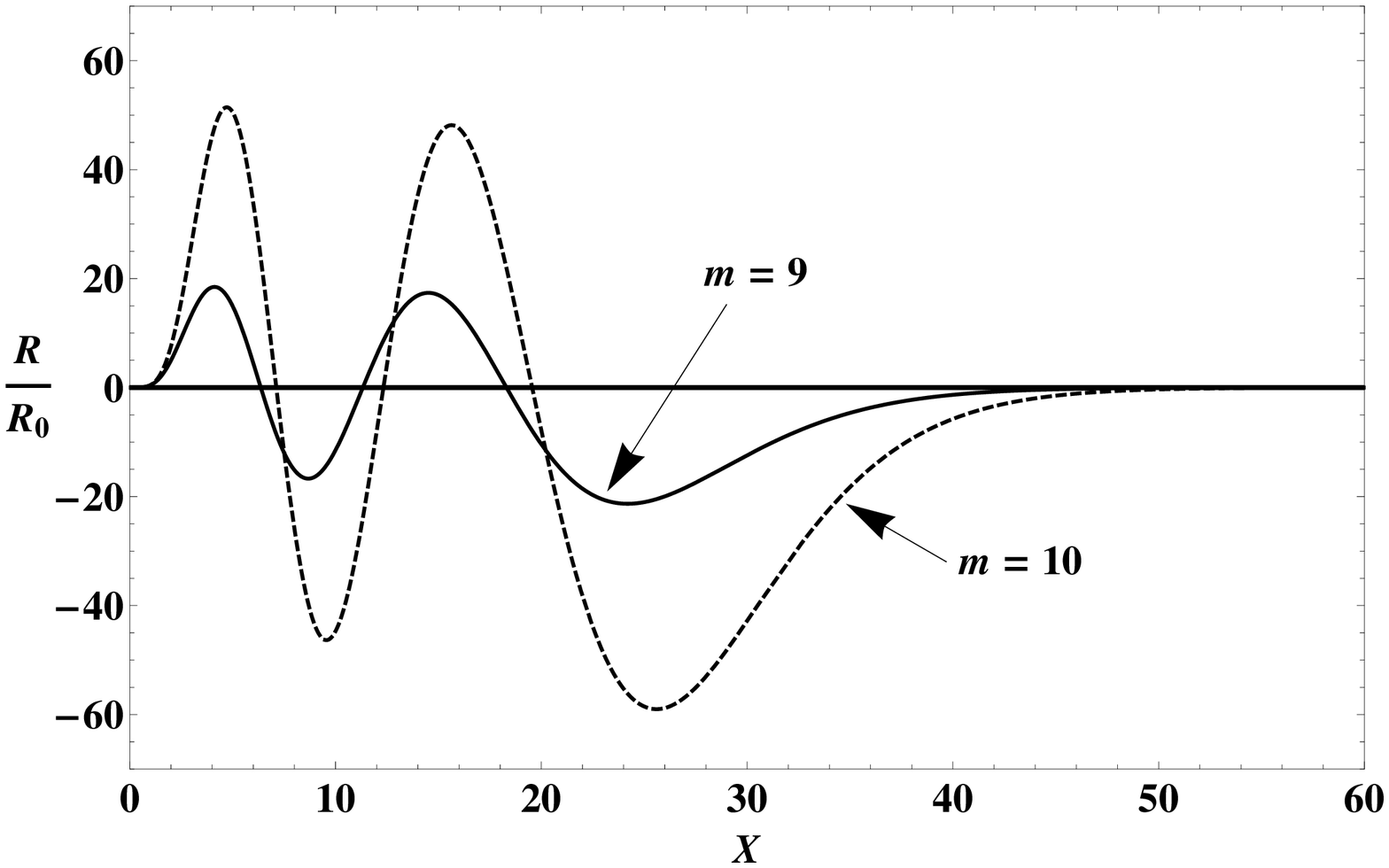}
\caption{\label{fig1}  {\sl Radial function $R$ as defined from Eq. (\ref{nrr}), in terms of $X = r^2/(2 \sigma^2)$. In the left panel, $l = 0, 1, 2$ for a fixed $m = 0$. In the right panel, $m = 9, 10$ for a fixed $l = 3$. We note that always $R(0) = 0$.}}
\end{figure}

In conclusion, the radial function is given by
\begin{equation}
R(r) = R_0 e^{-X/2}\,X^{|m|/2 + 1/4}\,{\cal M}(-l, 1+|m|, X) \,,
\label{nrr}
\end{equation}
where $R_0$ is a normalization constant. Equations (\ref{norma}), (\ref{sca}) and (\ref{nrr})  give
\begin{equation}
\label{no}
R_{0}^2 = - \frac{\sqrt{2}\,M\,c^2}{\Delta\,\sigma\,q\,\phi_{0}}\left[\int_{0}^{\infty}dX\,\left(1 + \frac{H^2\,X}{2}\right)^{1/2}\,e^{-X}\,X^{|m|}\,\Bigl({\cal M}(- l, 1 + |m|, X)\Bigr)^2\right]^{-1} \,.
\end{equation}
The integral on the right-hand side of Eq. (\ref{no}) can be numerically obtained for specific values of $H, m, l$. For consistency, $R_{0}^2 > 0$ imply $q\,\phi_0 < 0$. The radial wave function is everywhere well-behaved, and has $l + 1$ nodes as apparent in Fig. \ref{fig1}.

From Eq. (\ref{ext}) we have
\begin{equation}
\label{exte}
\rho^{\rm ext} = \rho_{0}^{\rm ext}\,\left(\frac{\phi_0}{\phi}\right)^3\,\left(1+\frac{H^2\,X}{4}\right) \,, \quad \rho_{0}^{\rm ext} = - \frac{\varepsilon_0\,\phi_0\,H^2}{2\,\sigma^2} \,,
\end{equation}
showing that the $\phi_0 > 0$ corresponds to a negative external charge density, and reciprocally.  Asymptotically, one has $\rho^{\rm ext} \sim 1/r$ for $r \gg \sigma$. Similar expressions can be found for the external current density.

\begin{figure}
\includegraphics[angle=0,scale=0.5]{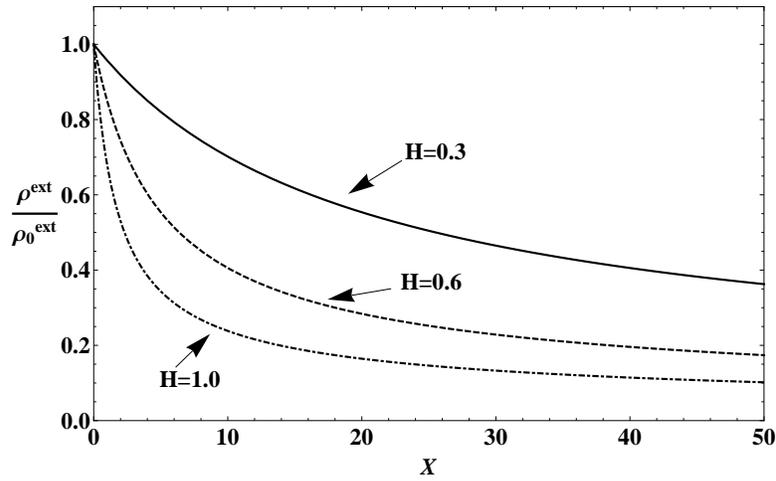}
\caption{\label{fig2}  {\sl External charge density from Eq. (\ref{exte}) as a function of $X = r^2/(2\,\sigma^2)$ for different values of the quantum diffraction parameter $H$ in 
Eq. (\ref{dif}). Upper curve (line): $H = 0.3$; middle curve (dashed): $H = 0.6$; lower curve (dot-dashed): $H = 1.0$.}}
\end{figure}

The external charge density is a monotonously decreasing function of position as seen in Fig. \ref{fig2}. On the other hand, the charge density associated to the test charge is found from Eq. (\ref{den}) to be
\begin{equation}
\label{fu}
\rho = \rho_0\,\sqrt{1+\frac{H^2\,X}{2}}\,e^{-X}\,X^{|m|}\,\Bigl({\cal M}(- l, 1 + |m|, X)\Bigr)^2\,\sin^{2}(n\,\theta) \,, \quad \rho_0 = - \frac{q^2\,\phi_0\,R_{0}^2}{\sqrt{2}\,\pi\,M\,c^2\,\sigma} \,.
\end{equation}
%

%
%

%
%

Finally, from Eq. (\ref{ford}) the confining force density on the test charge in the example is
\begin{equation}
\label{fd}
\rho\,{\bf E} + {\bf J}\times{\bf B} = - \frac{\hbar^2}{4\,M\,\sigma^4}\,|\Psi|^2\,r\,\hat{r} \,.
\end{equation}
Although the effective photon mass does not explicitly appear in Eq. (\ref{fd}), it plays a role in several steps of the derivation. For instance, the EM field in Eq. (\ref{eb}) becomes singular if $m_{\rm ph} \rightarrow 0$.

\subsection{Radial electric field and azimuthal magnetic field of constant strengths}
\label{3b}
Supposing a linear scalar potential
\begin{equation}
\label{phi}
\phi = - E_0 r \,,
\end{equation}
where $E_0$ is a constant, from Eq. (\ref{emf}) one has the radial electric field ${\bf E} = E_0 \hat{r}$, 
and the azimuthal magnetic field ${\bf B} = [E_0 \omega/(c^2 k)] \hat{\varphi}$, both of constant strength. This configuration provides a confinement in the radial direction.

Defining the new variable
\begin{equation}
X = \frac{m_{\rm ph} |q E_0| r^2}{\hbar^2 k}
\end{equation}
and the transformation
\begin{equation}
R = e^{X/2} X^{|m|/2+1/4} \Phi(X)
\end{equation}
the result from Eq. (\ref{r}) is
\begin{equation}
X\, \frac{d^2\Phi}{dX^2} + \left(1 + |m| + X\right)\,\frac{d\Phi}{dX} + \frac{1}{2}\left(\frac{k (P_{0}^2 - M^2 c^2)}{2 m_{\rm ph} |q E_0|} + 1 + |m|\right) \Phi = 0 \,,
\end{equation}
which is a Kummer equation also, identical to Eq. (\ref{ku}) after the replacement $X \rightarrow - X$. Proceeding as in the last subsection, one derives the regular solution
\begin{equation}
R = R_0 e^{X/2} X^{|m|/2+1/4} {\cal M}(1 + |m| + l, 1 + |m|, - X) \,,
\end{equation}
where ${\cal M}(1 + |m| + l, 1 + |m|, - X)$ is the Kummer confluent hypergeometric function of the indicated arguments and where the quantization condition
\begin{equation}
P_{0}^2 = M^2 c^2 +  \frac{2 m_{\rm ph} |q E_0|}{k}  \,(1 + |m| + 2 l) \,, \quad l = 0, 1, 2,...
\end{equation}
holds. 

In addition, working as in the last example we find the normalization constant
\begin{equation}
R_{0}^2 = \frac{2 m_{\rm ph} M c^2}{\hbar^2 k \Delta}\left[\int_{0}^{\infty} dX e^X X^{|m|+1/2}\Bigl({\cal M}(1 + |m| + l, 1 + |m|, - X)\Bigr)^2\right]^{-1} \,,
\end{equation}
the external charge density
\begin{equation}
\rho^{\rm ext} = \varepsilon_0 E_0/r \,,
\end{equation}
the test particle charge density
\begin{equation}
\rho = \rho_0 \,e^X X^{|m|+1/2} \Bigl({\cal M}(1 + |m| + l, 1 + |m|, - X)\Bigr)^2 \sin^{2}(n\theta)  \,, \quad \rho_0 = q |q E_0| R_{0}^2/(\pi M c^2) \,,
\end{equation}
and the force density
\begin{equation}
\rho{\bf E} + {\bf J}\times{\bf B} = - \frac{1}{M}\left(\frac{q E_0 m_{\rm ph}}{\hbar k}\right)^2 |\Psi|^2 r \hat{r} \,.
\end{equation}

\section{Conservation laws, stability analysis and numerical results}

In this Section we investigate the stability of the solutions found, by direct comparison with the numerical simulation of the KGE. For the validation of the simulations, it is important to verify the conservation laws $dQ/dt = 0, d{\cal H}/dt = 0$, where
\begin{eqnarray}
\label{q}
Q &=& \frac{q}{2\,M\,c^2}\int d{\bf r}\left[i\hbar\left(\Psi^{*}\frac{\partial\Psi}{\partial t} - \Psi\frac{\partial\Psi^*}{\partial t}\right) - 2\,(q\,\phi + {\cal E})\,|\Psi|^2\right] \,, \\
{\cal H} = \frac{1}{4\,M}\int d{\bf r}\Bigl[\frac{\hbar^2}{c^2}\frac{\partial\Psi^*}{\partial\,t}\,\frac{\partial\Psi}{\partial\,t} &+& \hbar^2\nabla\Psi^{*}\cdot\nabla\Psi + \Bigl(M^2\,c^2 - (q\,A^{\mu} + p^\mu)(q\,A_\mu+p_\mu)\Bigr)\,|\Psi|^2 \nonumber \\ &+& i\,\hbar\,(q{\bf A}+{\bf p})\cdot\Bigl(\Psi^{*}\nabla\Psi - \Psi\nabla\Psi^*\Bigr)\Bigr]
\label{h}
\end{eqnarray}
are, respectively, the total charge and Hamiltonian functionals associated to the test charge, where $\Psi$ satisfy Eq. (\ref{aa2}), and where $A^\mu$ in Eqs. (\ref{q}) and (\ref{h}) is the shifted four-potential according to Eq. (\ref{shi}). 
These conservation laws are a consequence of the Noether invariance of the action functional
\begin{equation}
\label{act}
S_{\rm act}[\Psi,\Psi^{*}] = \int\,d^4 x\,{\cal L} \,, \quad {\cal L} = \frac{1}{4\,M}\Bigl(\hbar\,\partial^{\mu}\Psi^* - i\,(q\,A^{\mu} + p^\mu)\Psi^{*}\Bigr)\,\Bigl(\hbar\,\partial_{\mu}\Psi + i\,(q\,A_{\mu}+p_\mu)\Psi\Bigr) - \frac{M\,c^2}{4}\,|\Psi|^2 
\end{equation}
under local gauge transformations and time translations (in our case $A_\mu$ is time-independent). It is a simple matter to show that the functional derivatives $\delta\,S_{\rm act}/\delta\Psi^{*} = 0$ and $\delta\,S_{\rm act}/\delta\Psi = 0$ generate Eqs. (\ref{aa2}) and its complex conjugate, respectively, and that the Legendre transform 
from Eq. (\ref{act}) produces the Hamiltonian (\ref{h}). 

For the exact solution of Sec. \ref{esol}, the charge conservation is equivalent to $Q = \int d{\bf r} \rho$, where the aforementioned test charge density $\rho$ is given by Eq. (\ref{den}). On the other hand, the energy conservation law (\ref{h}) for ${\bf p} = 0$ explicitly reads 
\begin{eqnarray}
{\cal H} &=& \frac{\Delta}{2\,M} \int_{0}^{\infty} dr \, \Bigl\{\hbar^2 \Bigl[\Bigl(\frac{dR}{dr}\Bigr)^2 - \frac{R}{r}\frac{dR}{dr}\Bigr] + \Bigl[M^2 c^2 + \frac{\hbar^2}{r^2}\Bigl(m^2 + \frac{1}{4}\Bigr) + \nonumber \\ &+& n^2\,\hbar^2 \Bigl(\frac{\omega^2}{c^2} + k^2\Bigr) - 2\,M\,q\phi + \Bigl(\frac{\omega^2}{c^2\,k^2} - 1\Bigr)\,\frac{q^2\,\phi^2}{c^2}\Bigr]\,R^2\Bigr\} \,.
\label{inter}
\end{eqnarray}
A few algebraic steps consider integrating by parts the first two terms in Eq. (\ref{inter}) assuming decaying boundary conditions, plus the use of the dispersion relation (\ref{a5}), 
the KGE (\ref{aa2}), the radial equation (\ref{r}), the  definition (\ref{ss}) and the normalization condition (\ref{norma}). In such way we finally derive the simple expression 
\begin{equation}
{\cal H} = M\,c^2 + \frac{\Delta\,n^2\,\hbar^2\,\omega^2}{M\,c^2}\int_{0}^{\infty}dr\,R^2 \,,
\label{hh}
\end{equation}
which is valid in this particular case. In Eq. (\ref{hh}) the second term $\sim \omega^2$ shows in a transparent way the contribution of the plasma wave to the energy. In a frame where the asymptotic momentum ${\bf p} \neq 0$ the form of ${\cal H}$ is a little more complicated due to coupling between translational and rotational degrees of freedom, and hence will be omitted.

For the numerical simulations, consider $q, M$ as the electron charge and mass and the solution in Sec. \ref{3a}. Rewrite the quantization condition (\ref{qua}) as
\begin{equation}
\label{xx}
n^2 = \frac{M^2}{m_{\rm ph}^2} + \frac{\hbar^2}{m_{\rm ph}^2 c^2 \sigma^2}\,(1+ |m| + 2 l) + \frac{q^2 \phi_{0}^2}{\hbar^2 k^2 c^2}  \,,
\end{equation}
where $n = 1, 2, 3,...$, $m = 0, \pm 1, \pm 2,...$, $l = 0, 1, 2,...$. 

Equation (\ref{xx}) has several free parameters. For definiteness, we chose the three terms on the right-hand side (respectively, proportional to $M^2, \hbar^2$ and $q^2$ to be of the same magnitude. This corresponds to similar contributions from the rest energy, the kinetic energy and the EM field energy. In this case, $n = \sqrt{3} M/m_{\rm ph}$. One might estimate \cite{Mendonca3, Akhiezer} the effective mass of transverse photons by the Akhiezer-Polovin relation $m_{\rm ph}\,c^2 = \hbar\,\omega_p$, where $\omega_p = \sqrt{n_0\,q^2/(M\varepsilon_0)}$ is the plasmon frequency and $n_0$ is the number density $n_0$. A more detailed, QED calculation of the photon mass in presence of a CPEM wave can be found in \cite{Raicher2}. For $n = 1000$, one finds $n_0 = 5.7 \times 10^{32} \,{\rm m}^{-3}$, which is in the limit of today's laser facilities \cite{Eliasson1, Eliasson2}. For  $n = 100$, one
has $n_0 = 5.7 \times 10^{34} \, {\rm m}^{-3}$, while $n = 10$ deserves $n_0 = 5.7 \times 10^{36} \,{\rm m}^{-3}$ (white dwarf).
Moreover, for $m = 0, l = 0$ one has $\sigma = \lambda_C/(2\pi) = 3.9 \times 10^{-13} \,{\rm m}$, where $\lambda_C = 2\pi\hbar/(M c)$ is the electron Compton length, besides a quantum diffraction parameter $H = 1$. Finally, to satisfy $M/m_{\rm ph} = |q \phi_0|/(\hbar k c)$ some free choices are still available. To avoid pair creation we set a not too large energy $|q\phi_0| = 0.1 M c^2 = 0.05 \,{\rm MeV}$ and calculate the wave-number $k$. The  results are shown in Table I, where the wavelength $\lambda = 2\pi/k$ and the angular frequency $\omega$ are also displayed. We find a range from the extreme ultraviolet to the hard X-ray radiation. Notice that it is not unusual to consider highly oscillating solutions to the KGE. For instance, consider the discussion of higher harmonic solutions of the KGE with a  large $n$, in the context of a charged particle propagation under strong laser fields in underdense plasmas \cite{Varro2}.  Possible experimental realization of the confining EM fields would involve high-intensity-laser-driven Z pinches as described in Ref. \cite{Beg}. As apparent from Eq. (\ref{ext}), necessarily a longitudinal external current should be set up, with the adequate radial dependence to fit the four-potential.


\begin{table}
\begin{tabular}{ |c|c|c|c|c|c|c| }
 \hline
 $n$ & $n_0 \,({\rm m}^{-3})$ & $m_{\rm ph}/M$ & $k \,({\rm m}^{-1})$ & $\lambda \,({\rm m})$ & $\omega \,({\rm rad/s})$  \\
\hline
10 & $5.70 \times 10^{36}$ & 0.173 & $4.49 \times 10^{10}$ & $1.40 \times 10^{-10}$ & $1.35 \times 10^{20}$  \\
\hline
100 & $5.70 \times 10^{34}$ & 0.017 & $4.49 \times 10^9$ & $1.40 \times 10^{-9}$ & $1.35 \times 10^{19}$ \\
\hline
1000 & $5.70 \times 10^{32}$ & 0.002 & $4.49 \times 10^8$ & $1.40 \times 10^{-8}$ & $1.35 \times 10^{18}$ \\
 \hline
\end{tabular}
\caption{Parameters for $m = 0, l = 0$ together with equal strength of the three terms on the right-hand side of the quantization
condition (\ref{xx}), for $|q\phi_0| = 0.1 M c^2$.}
\label{tab1}
\end{table}

For definiteness, choosing a frame where the test charge is at rest at infinity, one has ${\bf p} = 0, {\cal E} = M c^2$, which is adopted in the following. In order to simulate the problem, we use  Spectral Numerical Methods to solved the KGE (\ref{aa2}) in four-dimensional space with the analytic solution given in Sec. \ref{3a} as initial condition. We used box lengths $L_x = L_y = 3$ in the $x$ and $y$ dimensions, both normalized to $\sqrt{2}\sigma$, $L_z = 5$  in the $z$ direction (where periodic boundary conditions apply), normalized to $1/k$. We take the conditions of table \ref{tab1}. The spatial derivatives were approximated with a Fourier spectral method, performed with an implicit-explicit time stepping scheme. The space was resolved with $100$ grid points in the $x$ and $y$ directions and with $200$ grid points in the $z$ direction, and the time step was taken to be $\Delta t = 10^{-6}$, where time is normalized to $\omega^{-1}$. 

\begin{figure}
\begin{center}
\includegraphics[angle=0,scale=0.37]{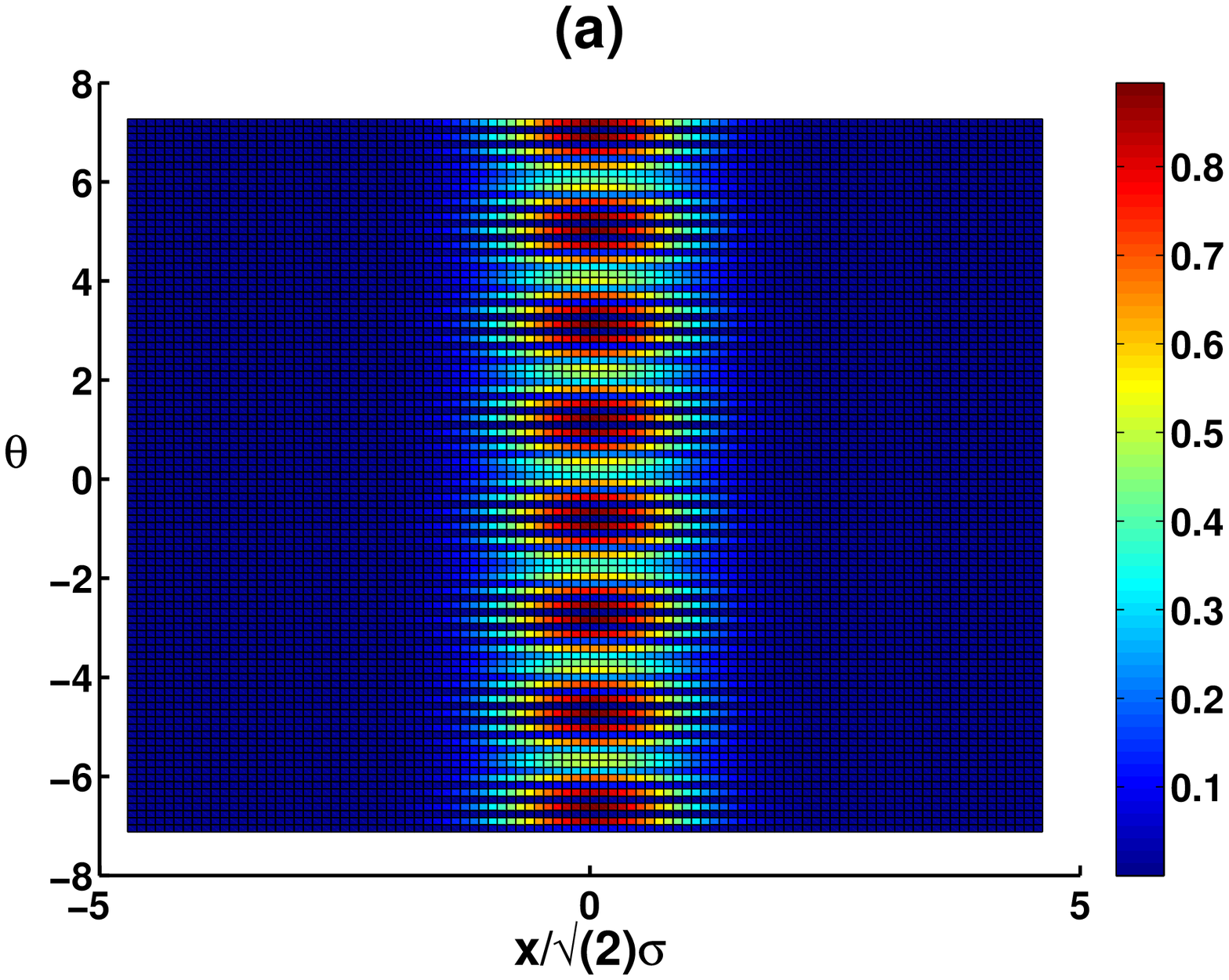}
\includegraphics[angle=0,scale=0.37]{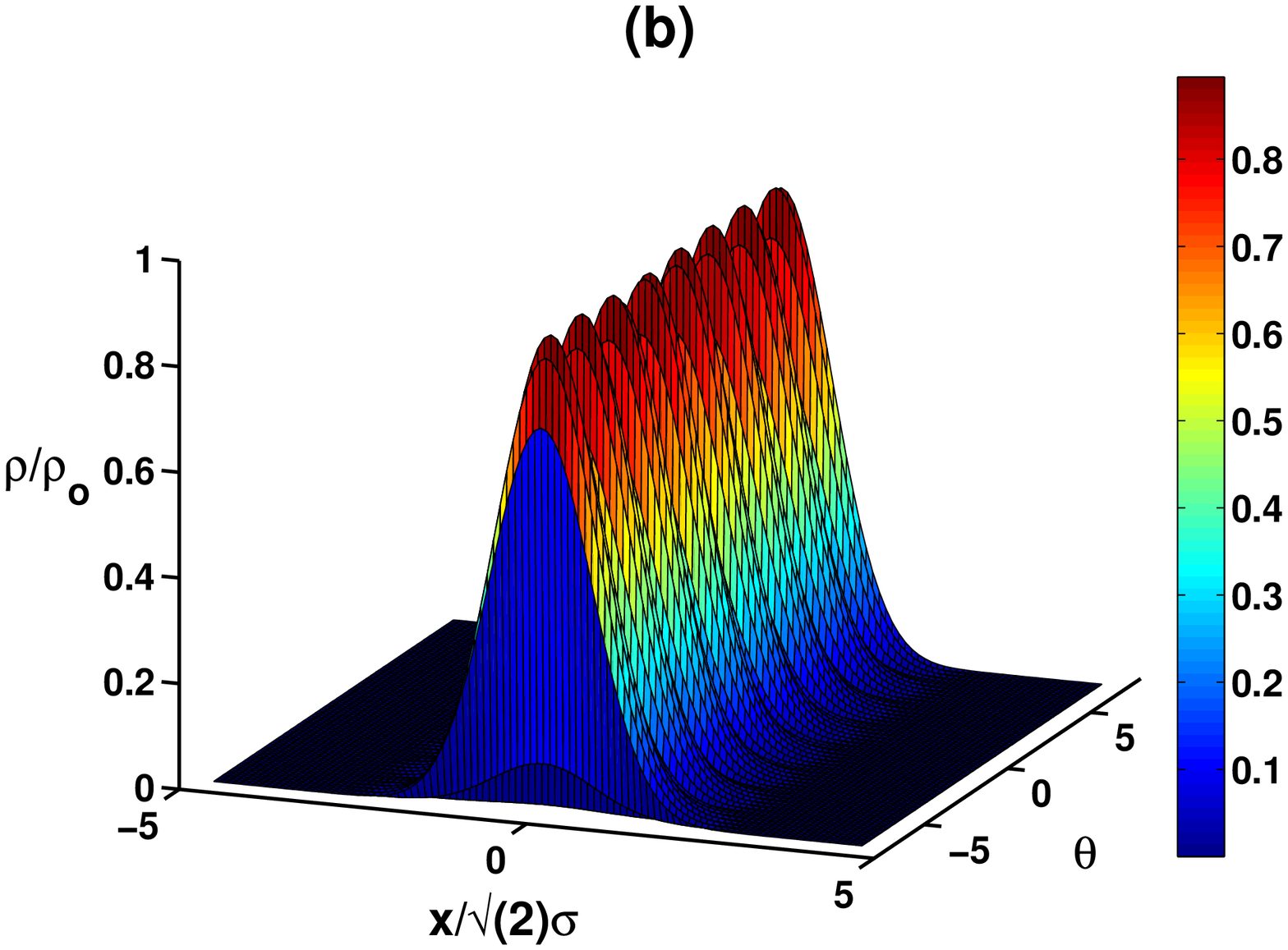}
\end{center}
\vspace{2.5cm}
\caption{{\sl Numerical simulation results for the charge density, obtained from the KGE (\ref{aa2}), in the $\theta-x$ plane at $y = 0$, for the states $m = 0, n = 10, l = 0$: 
(a) two-dimensional; (b) three-dimensional.}}
\label{fig4} 
\end{figure}

\begin{figure}
\begin{center}
\includegraphics[angle=0,scale=0.37]{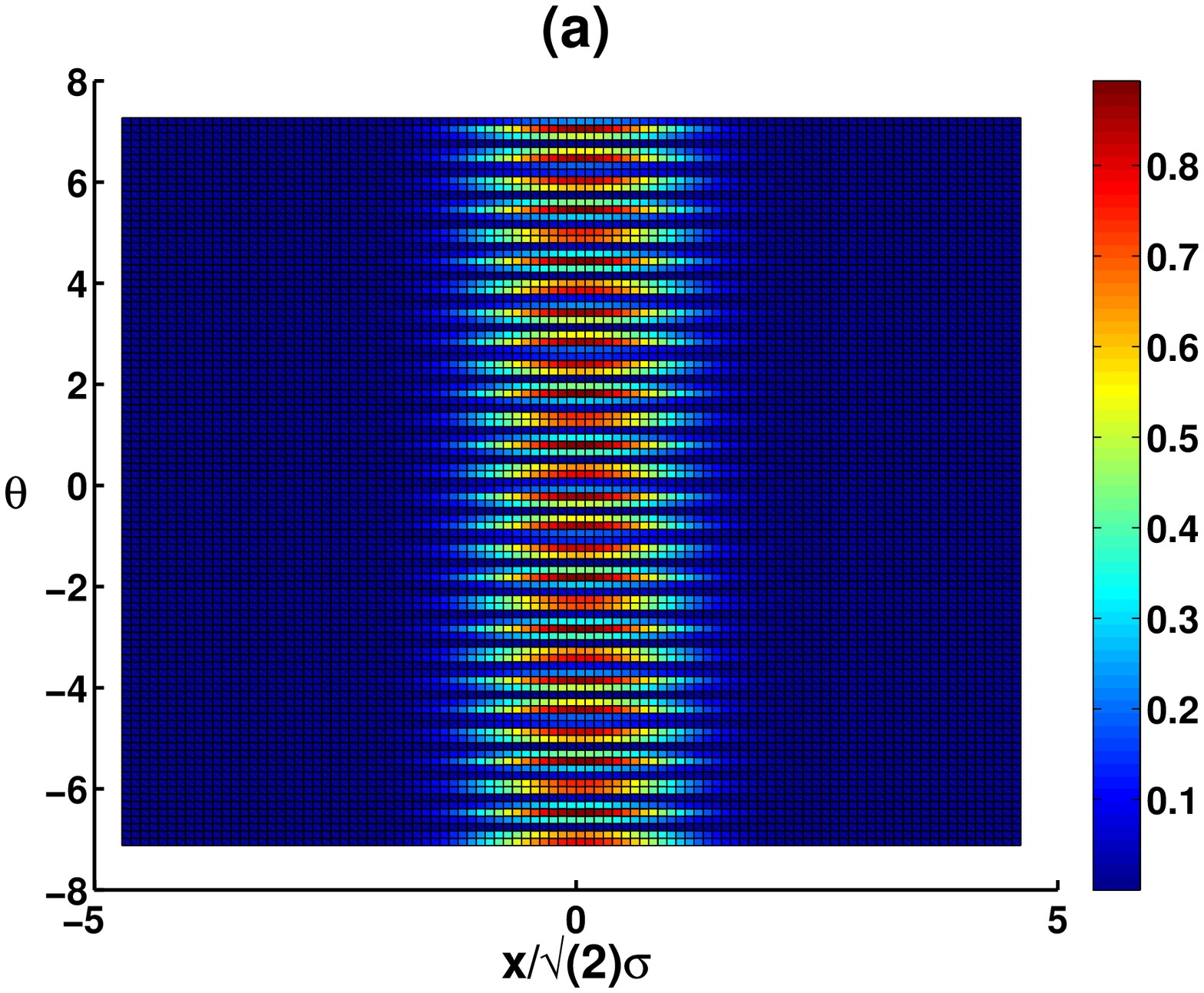}
\includegraphics[angle=0,scale=0.37]{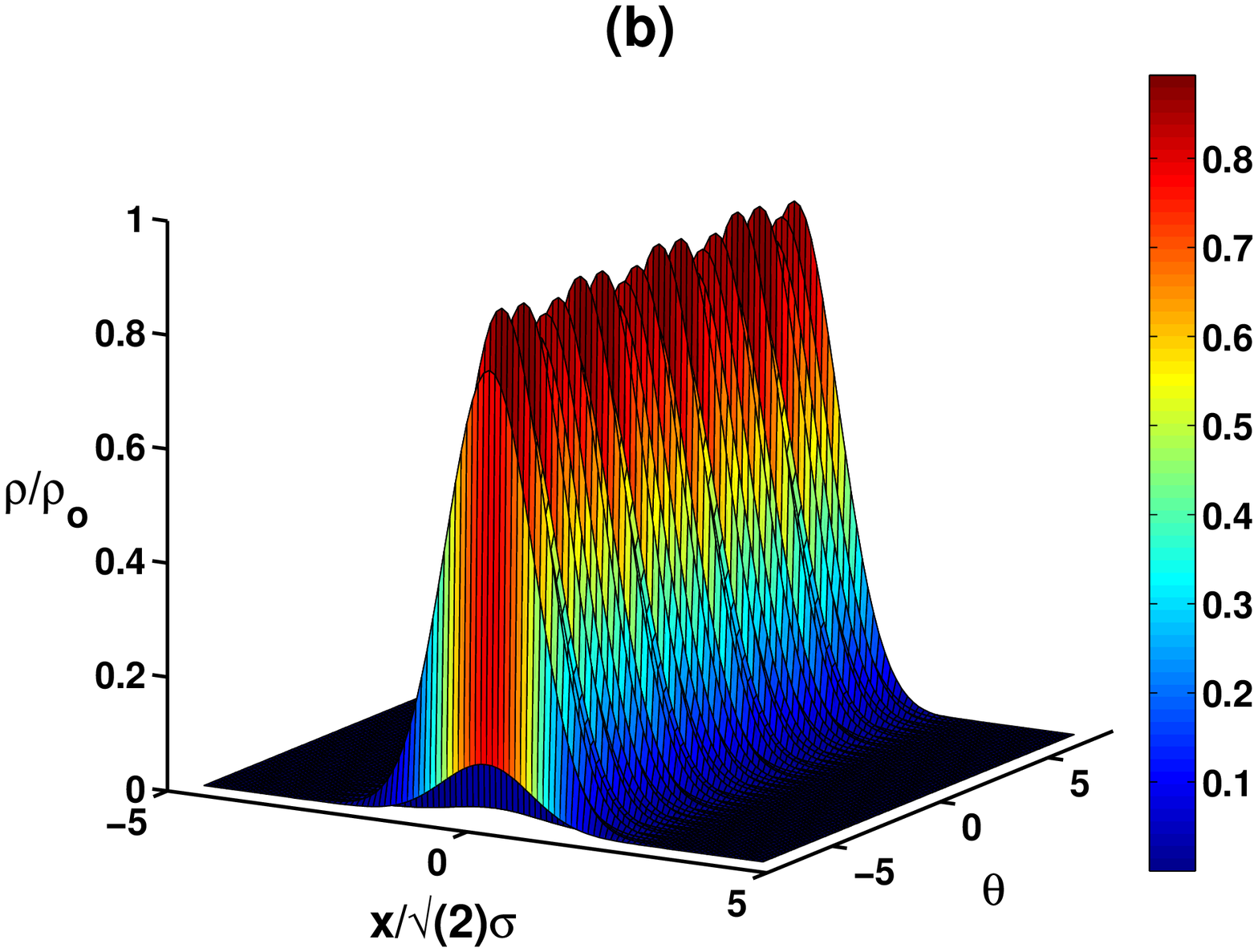}
\end{center}
\vspace{2.5cm}
\caption{\label{fig6}  {\sl Numerical simulation results for the charge density, obtained from the KGE (\ref{aa2}), in the $\theta-x$ plane at $y = 0$, for the states $m = 0, n = 100
0, l = 0$: (a) two-dimensional; (b) three-dimensional.}}
\end{figure}

In Figs. \ref{fig4} and \ref{fig6}, we have plotted the numerical result of the charge density of the test particle for $y = 0$, as a function of $x/(\sqrt{2}\sigma)$ and $\theta$, for the case of interest shown in the table \ref{tab1}, namely, $n=10, 100, 1000$. We used the parameters $m=0$, $\alpha=l=0$ and $H = 1$ showing an increase in the oscillation periods for  rising $n$.

%


%
%

%

\begin{figure}
\begin{center}
\includegraphics[angle=0,scale=0.37]{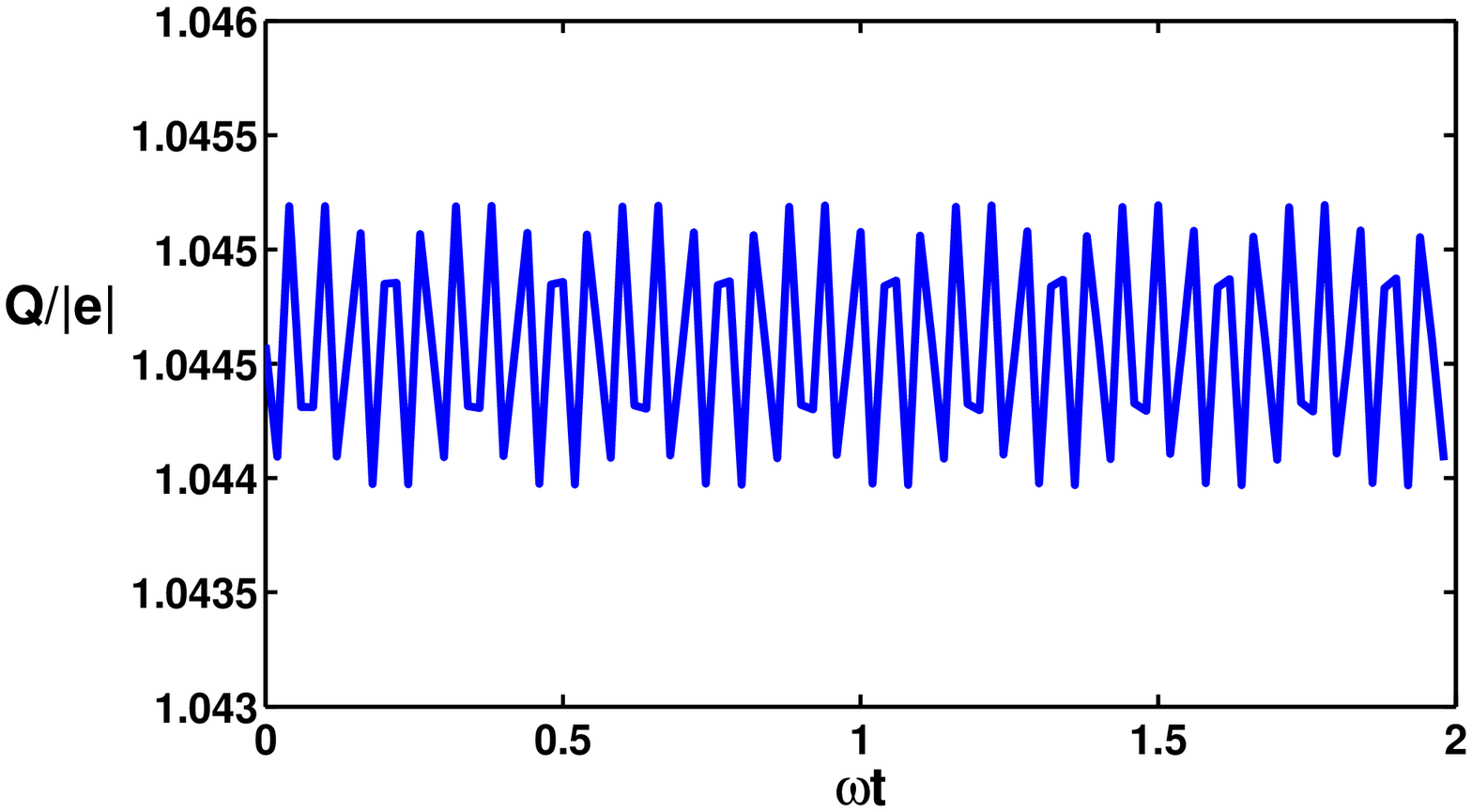}
\includegraphics[angle=0,scale=0.37]{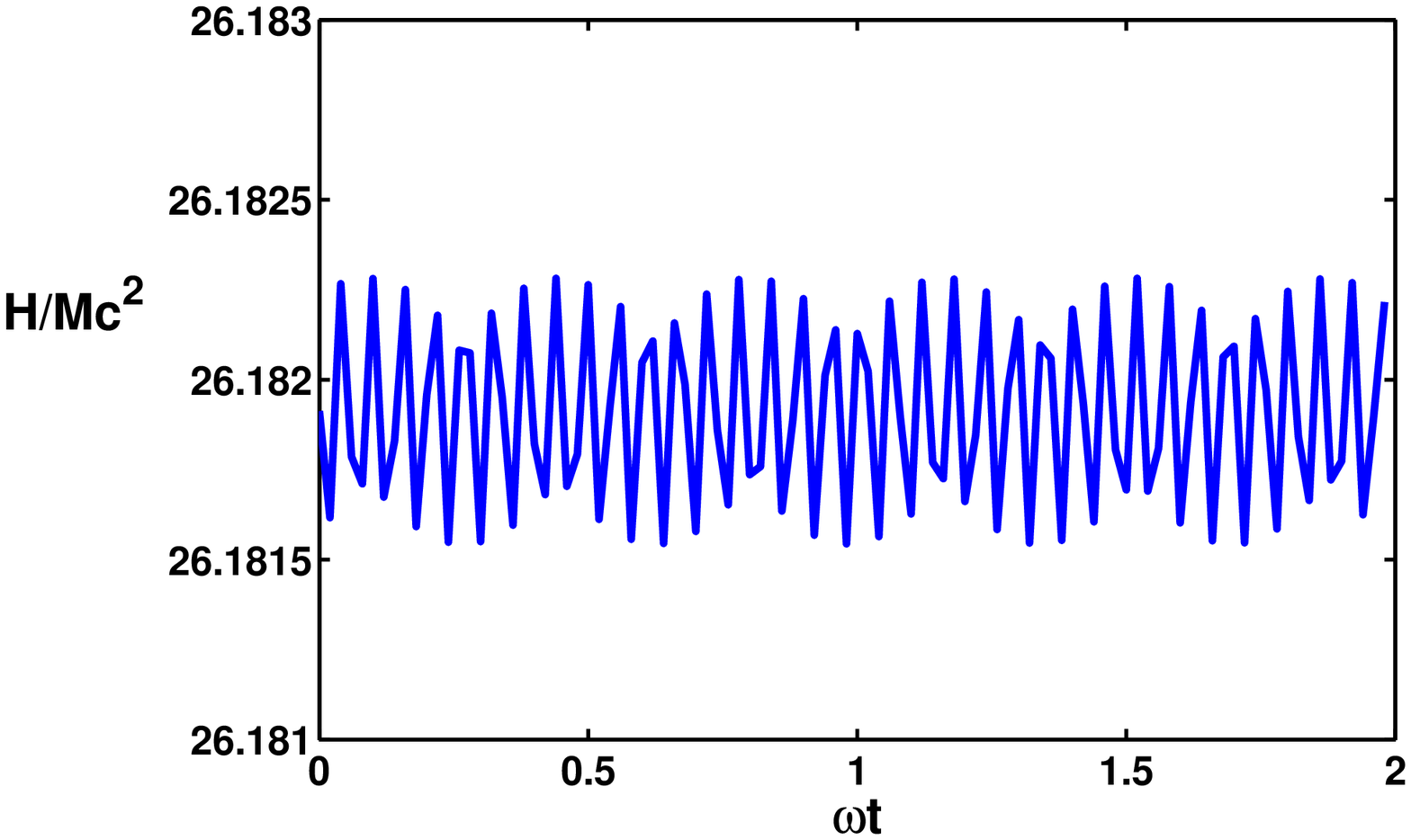}
\end{center}
\vspace{2.5cm}
\caption{\label{fig7}  {\sl Left: time-evolution of the global charge ${\cal Q}$ in Eq. (\ref{q}), normalized to the elementary charge $|e|$, for the state $(m=0, n=10, l=0)$. Right: time-evolution of the global energy ${\cal H}$ in Eq. (\ref{h}), normalized to $Mc^2$,
for the state $(m=0, n=10, l=0)$.}}
\end{figure}

To validate the simulations, the conservation laws of charge and total energy (\ref{q}) and (\ref{h}) were verified, as shown in Fig. 
\ref{fig7}. Fluctuations are small and differ from the exacts values in about $5\%$ for the state $(m=0, n=10, l=0)$.

\begin{figure}
\begin{center}
\includegraphics[angle=0,scale=0.37]{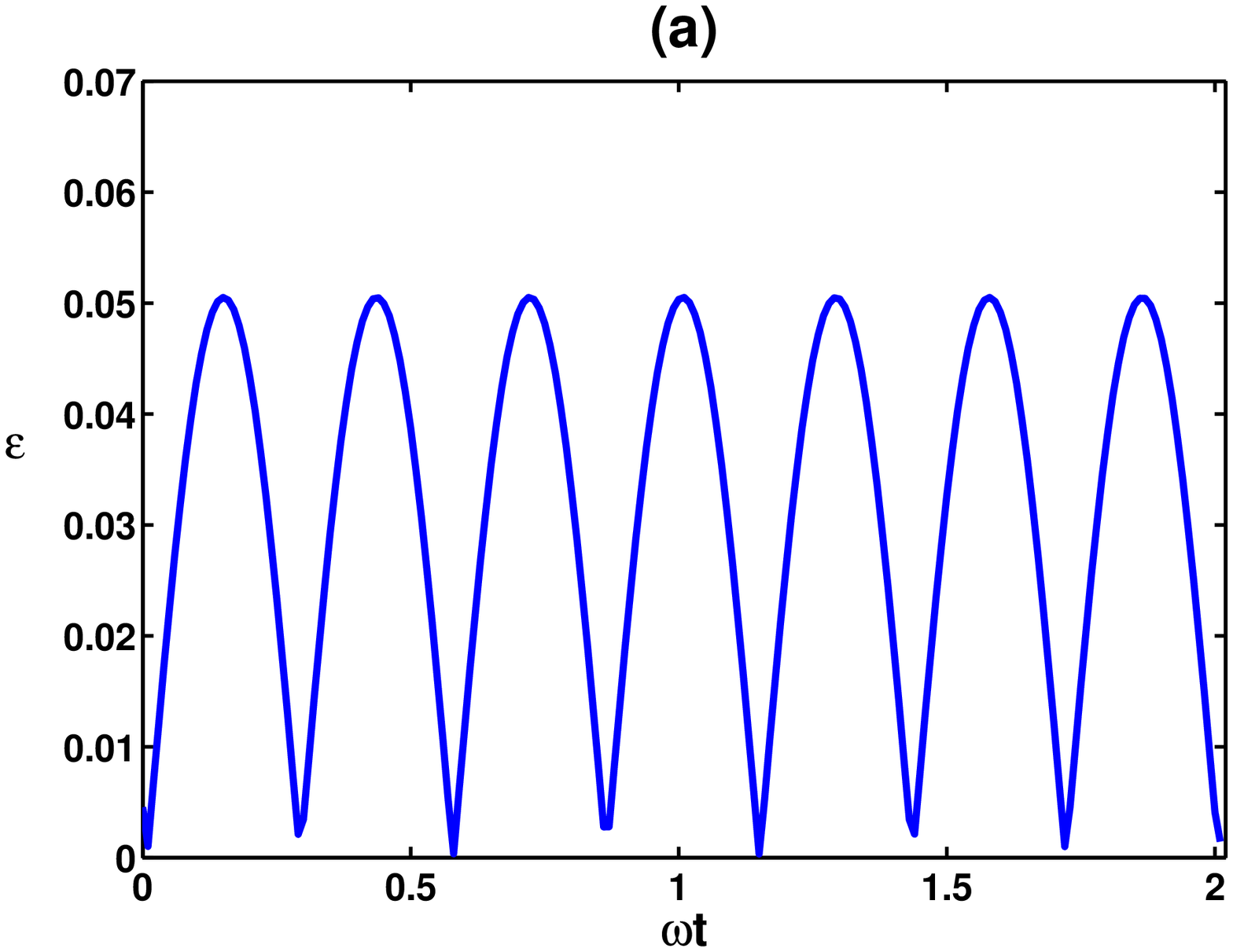}
\includegraphics[angle=0,scale=0.37]{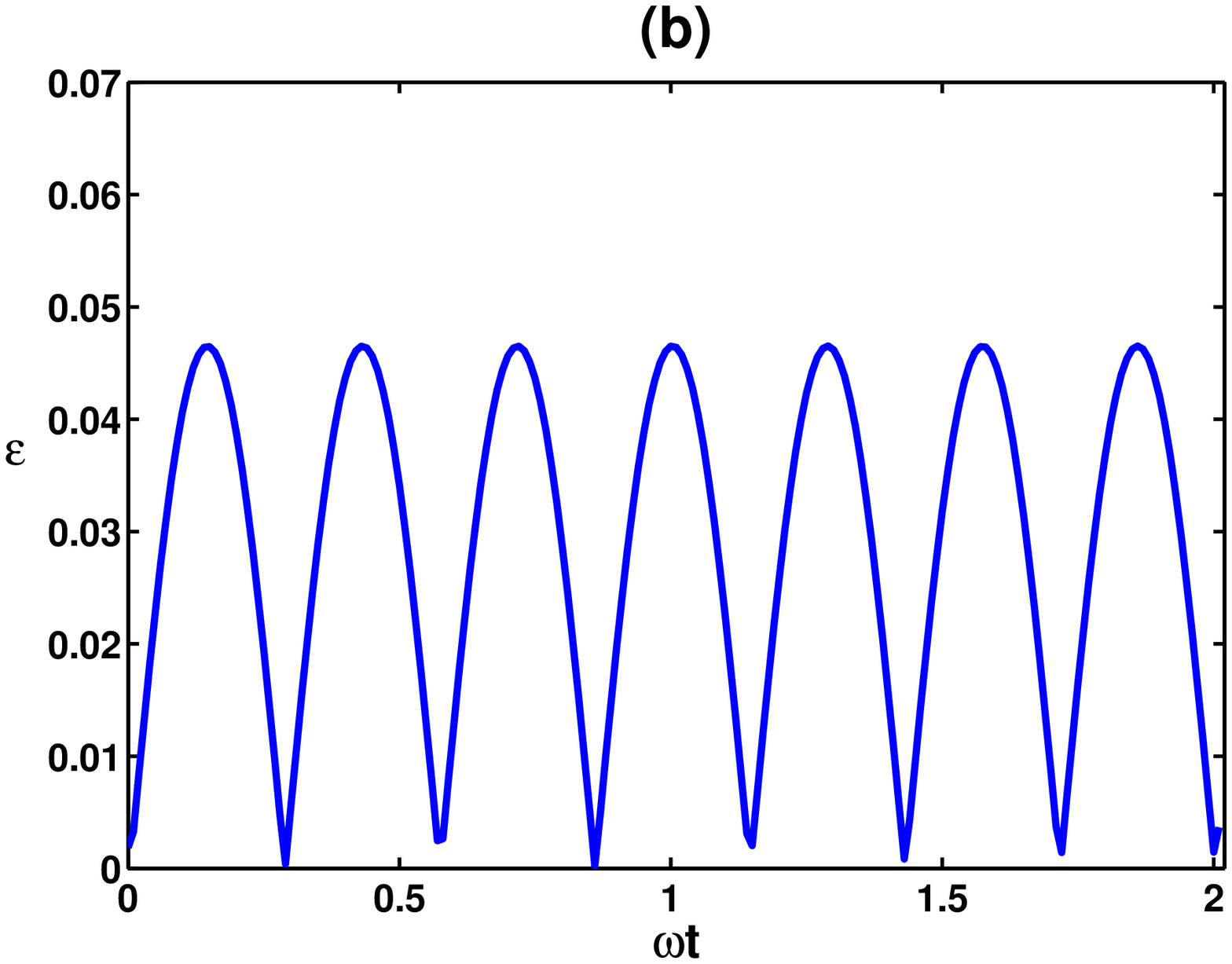}
\end{center}
\vspace{1.5cm}
\caption{\sl Relative deviation of the numerical solution from the exact analytic solution for random phase perturbations of
the exact state $(m=0, n=10, l=0)$. (a) Phase variation of $0.1$ rad; (b) phase variation of $0.05$ rad.}
\label{fig8}
\end{figure}

To numerically check the stability of the exact solution, we added random perturbations to the phase $\theta$ calculated at $t = 0$,  with aleatory angles between $0.1$ rad and $0.05$ rad. Figure \ref{fig8} shows the maximum relative error in charge density fluctuations $\varepsilon =|(\rho - \rho_{\rm num})|_{\rm max}/\rho_{\rm max}$, where $\rho$ follows from the analytical result in Sec. \ref{3a}, $\rho_{\rm num}$ is the numerical solution and $\rho_{\rm max}$ is the maximum value of the charge density analytically calculated, as a function of time, for the state $(m=0, n=10, l=0)$. 

\begin{figure}
\begin{center}
\includegraphics[angle=0,scale=0.37]{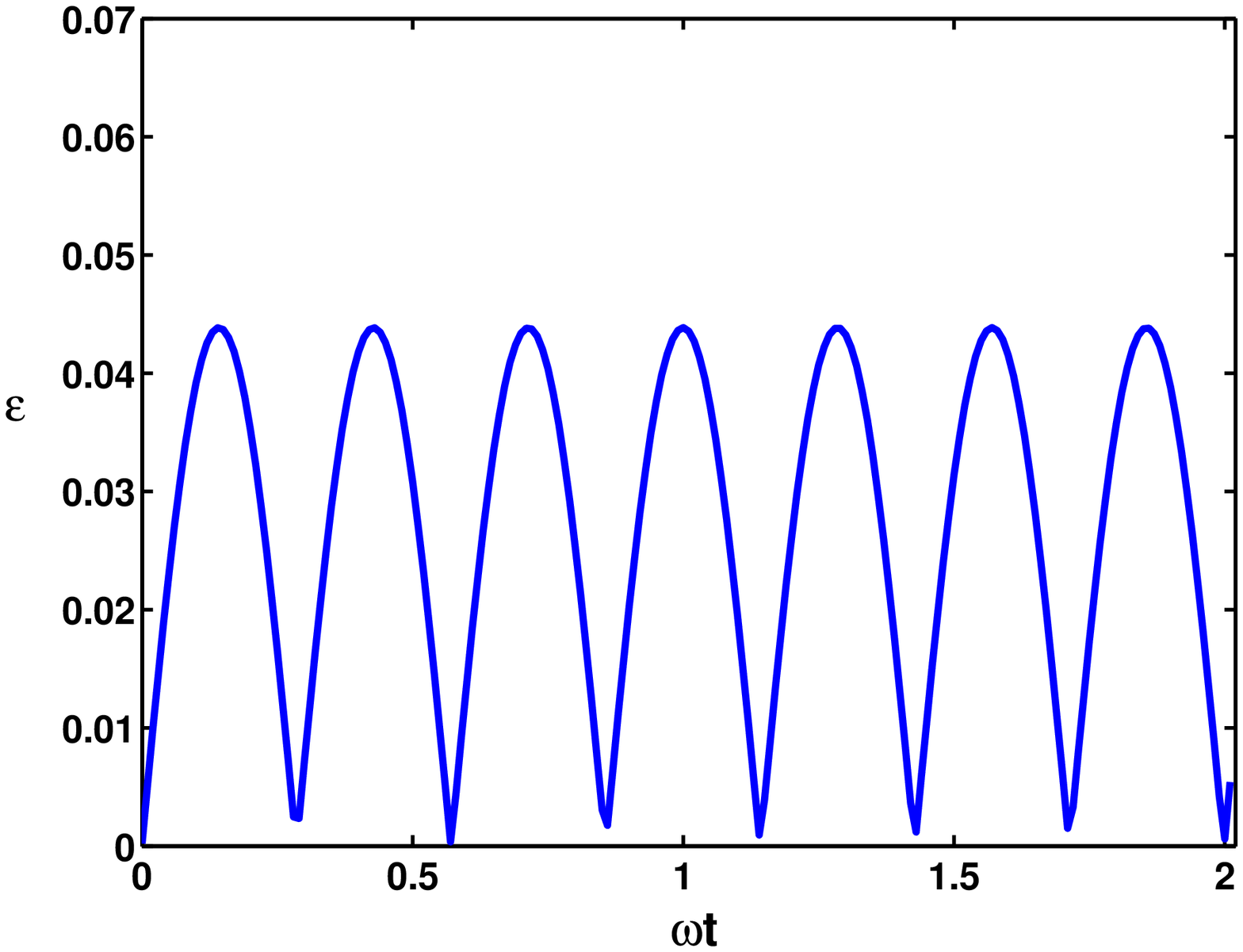}
\end{center}
\vspace{1.5cm}
\caption{\label{fig9}{\sl Relative deviation of the numerical solution from the exact analytic solution for random phase perturbations of
the exact state $(m=0, n=100, l=0)$ for a phase variation of $0.1$ rad.}}
\end{figure}

Similarly, Fig. \ref{fig9} shows the maximum relative error in charge density fluctuations for the state $(m=0, n=100, l=0)$.
The numerical solution almost exactly follows the analytic solution, without substantial changes throughout the simulation. For the case of the states
$(m=0, n=10, 100, 1000, l=1)$ there is a $5 \%$ relative error with stable oscillatory behavior. This result is maintained for different values of the random perturbations. Hence the compressed structures seems to be stable enough to be observable in experiments at least in the cases studied. Similar conclusions hold for the example of Sec. \ref{3b}.

In order to substantiate the numerical results, we also perform an analytical stability check, as follows. Assuming a phase perturbation according to
\begin{equation}
\label{pert}
\Psi = \exp\left(-\,\frac{i p \cdot x}{\hbar}\right)\,\frac{e^{im\varphi}}{\sqrt{r}} R(r) S(\theta + \delta\theta) \,, 
\end{equation}
plugging into Eq. (\ref{aa2}), where $R(r)$ and $S(\theta)$ satisfy Eqs. (\ref{s}) and (\ref{r}) with a four potential given by (\ref{4p}), and linearizing for $\delta\theta = \delta\theta(r,\varphi,z,t)$, gives a large equation which we refrain to show here.  To maintain the generality, the coefficients of $dR/dr$, $dS/d\theta$ and $S$ should vanish in this equation, otherwise only certain specific solutions for Eqs. (\ref{s}) and (\ref{r}) would be selected. We also adopt a reference frame where ${\bf p} = 0$. For an arbitrary scalar potential $\phi(r)$ and after some simple algebra, it can be shown that $\delta\theta = \delta\theta(\varphi)$, satisfying
\begin{equation}
\frac{d^2\delta\theta}{d\varphi^2} + 2\,i\,m\frac{d\delta\theta}{d\varphi} = 0 \,,
\end{equation}
possessing oscillatory solutions of the form $\delta\theta = c_0 + c_1\exp(-2\,i\,m\,\varphi)$ for constants $c_0, c_1$. The conclusion is that in this case we have linearly stable solutions. It should be noted that the restricted form of the perturbation (\ref{pert}) and the associated consistency analysis make the findings somehow limited. A full analytical stability check is beyond the scope of the present work.

\section{Conclusion}
In this work it was obtained a new exact solution for a charged scalar test charge.  
As an alternative to the traditional Volkov assumption, the quantum state contains a stringent dependence on the radial coordinate, mediated by the scalar 
potential $\phi(r)$ appearing in the fundamental equation (\ref{r}). The procedure can work only in a plasma medium, which implies a non-zero photon mass. However, by definition the setting is not of a quantum plasma, but of a quantum relativistic test charge under a classical plasma wave.  As discussed in Sec. III, for specific scalar potentials a quantization condition results from the requirement of a well-behaved radial wave function. The stability analysis of the solutions was numerically investigated by means of spectral methods. In a sense, the approach is complementary to the CPEM case, which assumes scalar and vector potentials respectively given by $\phi = 0, {\bf A} = {\bf A}_\perp$, as shown in Eq. (\ref{a3}), while in the present work $\phi \neq 0, {\bf A}_\perp = 0$. Applications for transverse compression in laser plasma interactions in the quantum relativistic regime or dense astrophysical settings with a high effective photon mass [or a  large $n$ compatible with the quantization condition (\ref{qua}), for instance] were discussed. The treatment is useful as a starting point for the collective, 
multi-particle coherent aspects of relativistic quantum plasmas. In this case, the EM field has to be calculated in a self-consistent way and not taken as an external input like in the present communication. Finally, the extension of the analysis to include spin is left to future work.

\acknowledgments

The authors acknowledge CNPq (Conselho Nacional de Desenvolvimento Cient\'{\i}fico e Tecnol\'ogico) for financial support and the anonymous Referee for the insightful 
constructive remarks.

\end{document}